\documentclass[preprint,prd,showpacs,amsmath,amssymb,amsthm,nofootinbib]{revtex4}
\usepackage[mathscr]{euscript}
\usepackage{bm}
\usepackage{graphicx}
\usepackage{subfigure}
\usepackage{overpic} 
\usepackage{multirow}  
\usepackage{floatrow}
\usepackage{array}
\usepackage{textcomp}
\usepackage[colorlinks=true,linkcolor=red]{hyperref}
\newfloatcommand{capbtabbox}{table}[][\FBwidth]
\newcommand{\bea}{\begin{eqnarray}}
\newcommand{\eea}{\end{eqnarray}}
\newcommand{\beq}{\begin{equation}}
\newcommand{\eeq}{\end{equation}}

\def\/{\over}

\begin{document}

\title{Primordial Black Holes from Inflation with Nonminimal Derivative Coupling }

\author{ Chengjie Fu,  Puxun Wu\footnote{Corresponding author: pxwu@hunnu.edu.cn} and Hongwei Yu }
\affiliation{Department of Physics and Synergetic Innovation Center for Quantum Effects and Applications, Hunan Normal University, Changsha, Hunan 410081, China}

\begin{abstract}
We propose a novel enhancement mechanism of the curvature perturbations in the nonminimal derivative coupling inflation model with a coupling parameter related to the inflaton field. By considering a special form of the coupling parameter as a function of the inflaton, a period of ultra-slow-roll inflation can be realized due to the gravitationally enhanced friction, and the resulting power spectrum of the curvature perturbations has a sharp peak, which is large enough to produce the primordial black holes. Under this mechanism, we can easily obtain a sharp mass spectrum of primordial black holes around specific masses such as $\mathcal{O}(10)M_\odot$, $\mathcal{O}(10^{-5})M_\odot$, and $\mathcal{O}(10^{-12})M_\odot$, which can explain the LIGO events, the ultrashort-timescale microlensing events in OGLE data, and the most of dark matter, respectively.

\end{abstract}


\maketitle
\section{Introduction}
\label{sec_in}
The idea of primordial black holes (PBHs) formed in advance of ordinary stars has been proposed for decades \cite{Hawking1971,Carr1974,Carr1975}, and these objects have long been taken as a potential candidate for dark matter (DM) \cite{Ivanov1994,Frampton2010,Khlopov2014}. Recently, they have been receiving renewed attention ever since the gravitational waves (GWs) were successfully detected. So far the LIGO-Virgo Collaboration has detected several events of GWs coming from the merger of black holes (BHs) \cite{GW1,GW2,GW3,GW4,GW5}. From these remarkable observations, it has been found that many of these  binary BHs have masses around $30M_\odot$ ($M_\odot$ is the solar mass) which are too heavy for BHs formed by stellar evolution. But such masses fall within the mass scales of PBHs, which can span many orders of magnitude contingent upon  their formation mechanism. Therefore, PBHs are considered as a promising candidate for the origin of BHs detected by the LIGO-Virgo Collaboration \cite{Bird2016,Clesse2017,Sasaki2016,Carr2016,Fernandez2019}.

In order to account for the LIGO-Virgo GW events by PBHs, it is required that PBHs with mass around $\mathcal{O}(10)M_\odot$ have a merger rate, $12-213\;\mathrm{Gpc}^{-3}\; \mathrm{yr}^{-1}$,  expected by the LIGO-Virgo Collaboration~\cite{GW3}. In Ref. \cite{Sasaki2016}, Sasaki \textit{et al.} have shown that PBHs comprising ${\mathcal{O}(1)}$\textperthousand\; of the total DM can realize such a merger rate. In addition to being taken as the GW source, PBHs could account for some microlensing events as well. Recently, it was pointed out that PBHs with a mass around $\mathcal{O}(10^{-5})M_\odot$, whose fraction in DM is of order $\mathcal{O}(10^{-2})$, can explain the ultrashort-timescale microlensing events in the OGLE data \cite{OGLE1,OGLE2}. Moreover,  recent analyses \cite{Katz2018,HSC} have shown that since the wave effect weakens the lensing effect, the gravitational femtolensing of gamma-ray bursts \cite{Femto} is not valid and the microlensing observation with the Subaru Hyper Supreme-Cam gives no constraint on PBHs below $10^{-11}M_\odot$. 
Since the observed distribution of white dwarfs can constrain the abundance of PBHs with mass around $10^{-14}-10^{-13}M_\odot$ \cite{WD}, so it is still possible to make up all DM by PBHs with mass around $10^{-16}-10^{-14}M_\odot$ and around $10^{-13}-10^{-11}M_\odot$. 
Thus, it is physically attractive to resolve the astrophysical and cosmological puzzles by PBHs.

The formation of PBHs requires a large primordial curvature perturbation produced during the inflationary phase. 
Recently, Cai \textit{et al.} \cite{Cai2018,Cai2019} suggested that the primordial curvature perturbations can be enhanced by the parametric resonance arising from the oscillating sound speed squared. And Refs. \cite{Ballesteros2019,Kamenshchik2019} have shown that a large amplification of curvature perturbations results from a sound speed approaching to zero during some stage of inflation in the single-field model with a non-canonical kinetic term. Moreover, in Ref. \cite{Pi2018}, the power spectrum of the curvature perturbations is amplified due to the oscillation during transition from scalaron dominated phase to the light field dominated phase in the model of Starobinsky's gravity theory with a nonminimally coupled scalar field. In addition, it has been found that  the quadruple-phase inflation can realize  three large peaks of the power spectrum at each phase transition~\cite{Tada2019}.
Of course, a more common mechanism of enhancing the curvature perturbations is the inflection-point inflation \cite{Bellido2017,Germani2017,Hu2017,Ezquiaga2018,Gong2018,
Ballesteros2018,Dalianis2019,Drees2019}, in which the potential has an approximate inflection point. Within this scenario, an inflaton experiences a very flat potential around the near-inflection point, and as a result, a  period of ultra-slow-roll inflation can be realized. Consequently, the curvature perturbations are amplified by many orders of magnitude due to the great decrease of the Hubble slow-roll parameter.   

A natural way to realize the ultra-slow-roll inflation is to flatten the potential, on one hand. But on the other hand,  there is also another feasible way to achieve the same goal, that is,  to slow down the evolution of inflaton by increasing friction. This  could be implemented by the mechanism of the gravitationally enhanced friction, which arises from a field derivative coupling with the Einstein tensor \cite{Germani2010,Tsujikawa2012}. In this paper, we will study the production of PBHs in the model of inflation with a nonminimal derivative coupling which results in enhanced friction. We organize our paper as follows: In Sec. \ref{sec2}, we outline the formulas about
the formation of PBHs. In Sec. \ref{sec3}, we discuss in detail how to produce the large-amplitude curvature perturbations during inflation in the framework of the nonminimal derivative coupling with enhanced friction. The results of the numerical calculation for the power spectrum of the curvature perturbations and the mass spectrum of PBHs are shown in Sec. \ref{sec4}. Finally, Sec. \ref{sec5} gives our conclusions and discussions.

\section{Formation of primordial black holes}
\label{sec2}
In this paper, we consider that the super-Hubble curvature perturbations produced in the phase of inflation reenter the horizon during the radiation-dominated era. If these perturbations are large enough so that the gravity of the overdense regions can overcome the radiation pressure, these regions will collapse to form PBHs soon after their horizon entry. The mass of formed PBHs is related to the horizon mass at the horizon entry of the perturbations with the comoving wave number $k$:
\begin{align}\label{M_PBH}
M(k)= \gamma \frac{4\pi }{\kappa^2H} \bigg|_{k=aH} \simeq M_\odot\left(\frac{\gamma}{0.2}\right)\left(\frac{g_\ast}{10.75}\right)^{-\frac{1}{6}}\left(\frac{k}{1.9\times10^6\;\mathrm{Mpc}^{-1}}\right)^{-2}\;,
\end{align}
where $\kappa^{-1}\equiv M_{\mathrm{pl}}=2.4\times10^{18}\;\mathrm{GeV}$ is the reduced Planck mass, $a$ is the cosmic scale factor, $H\equiv\dot{a}/a$ is the Hubble parameter and a dot denotes the derivative with respect to the cosmic time. $\gamma$ is the ratio of the PBH mass to the horizon mass and indicates the efficiency of collapse. Since $\gamma$ depends on the detail of gravitational collapse, the value of $\gamma$ has uncertainties. In our paper, we consider the case of $\gamma\simeq (1/\sqrt{3})^3$ estimated by the simple analytical calculation \cite{Carr1975}. $g_\ast$ denotes the effective degrees of freedom in the energy densities at the PBH formation. We consider that the PBHs are formed deep in the radiation-dominated era, and thus adopt $g_\ast=106.75$.

Under the assumption that the probability distribution function of perturbations is Gaussian, the production rate of PBHs with mass $M(k)$ based on the Press-Schechter theory is given by \cite{Young2014,Tada2019}
\begin{align}
\beta(M)=\int_{\delta_c}\frac{d\delta}{\sqrt{2\pi\sigma^2(M)}}e^{-\frac{\delta^2}{2\sigma^2(M)}}=\frac{1}{2}\mathrm{erfc}\left(\frac{\delta_c}{\sqrt{2\sigma^2(M)}}\right)\;,
\end{align}
where erfc denotes the complementary error function and $\delta_c$ is the threshold of the density perturbations for the PBH formation.\footnote{The calculations of $\beta(M)$ based on the peak theory were discussed  in detail  in Ref. \cite{Germani2019}.} Recent studies in Refs. \cite{Musco2013,Harada2013} have suggested the threshold value to be $\delta_c\simeq0.4$, and we adopt this value in the subsequent numerical calculations. Here the variance $\sigma^2(M)$ represents the coarse-grained density contrast with the smoothing scale $k$, which is defined as \cite{Young2014}
\begin{align}
\sigma^2(M(k))=\int d\ln{q} \ W^2(qk^{-1})\frac{16}{81}(qk^{-1})^4\mathcal{P_R}(q)\;,
\end{align}
where $\mathcal{P_R}(k)$ is the power spectrum of the primordial curvature perturbations, and $W$ is the window function, which, in our analysis, is  taken to be  the Gaussian function $W(x)=e^{-x^2/2}$.

The current fraction of PBHs against the total DM is given by
\begin{align}
\frac{\Omega_{\mathrm{PBH}}}{\Omega_{\mathrm{DM}}}= \int \frac{dM}{M} f(M)\;,
\end{align}
where
\begin{align}\label{f_pbh}
f(M)\equiv \frac{1}{\Omega_{\mathrm{DM}}} \frac{d\Omega_{\mathrm{PBH}}}{d\ln{M}} 
\simeq \frac{\beta(M)}{1.84\times10^{-8}}\left(\frac{\gamma}{0.2}\right)^{\frac{3}{2}}\left(\frac{10.75}{g_\ast}\right)^{\frac{1}{4}}\left(\frac{0.12}{\Omega_{\mathrm{DM}}h^2}\right)\left(\frac{M}{M_\odot}\right)^{-\frac{1}{2}}\;.
\end{align}
$\Omega_{\mathrm{DM}}$ is the current density parameter of DM, and the Planck 2018 results \cite{1Planck2018} give its value $\Omega_{\mathrm{DM}}h^2\simeq0.12$. From above equations, one can note that the production of a sizeable amount of PBHs requires that the typical curvature perturbations are significant $\mathcal{P_R}\sim \mathcal{O}(10^{-2})$ on the small scales, which is about seven orders of magnitude larger than the perturbations with $\mathcal{P_R}\sim \mathcal{O}(10^{-9})$ on the the cosmic microwave background (CMB) scales. 

In this paper, we focus on the PBHs with mass around $\mathcal{O}(10)M_\odot$, $\mathcal{O}(10^{-5})M_\odot$, and $\mathcal{O}(10^{-12})M_\odot$, which are mentioned in the Introduction. From Eq. (\ref{M_PBH}), we can estimate the scales of the curvature perturbations corresponding these PBH masses to be $\mathcal{O}(10^{5})\mathrm{Mpc}^{-1}$, $\mathcal{O}(10^{8})\mathrm{Mpc}^{-1}$, and $\mathcal{O}(10^{12})\mathrm{Mpc}^{-1}$, respectively. We expect that the power spectrum of the curvature perturbations have a large peak at these scales.

\section{Nonminimal derivative coupling model}
\label{sec3}
In this section, we will qualitatively investigate how to achieve an amplification of curvature perturbations by the mechanism of the gravitationally enhanced friction in the framework of the nonminimal derivative coupling between the gravity and the inflaton field $\phi$. The action of the nonminimal derivative coupling model has the form
\begin{align}\label{action}
\mathcal{S}=\int d^4x \sqrt{-g}\left[\frac{1}{2\kappa^2}R - \frac{1}{2}\left(g^{\mu\nu}-\kappa^2 \xi G^{\mu\nu}\right)\nabla_\mu\phi\nabla_\nu\phi - V(\phi)\right]\;,
\end{align}
where $g$ is the determinant of the metric tensor $g_{\mu\nu}$, $R$ is the Ricci scalar, $G^{\mu\nu}$ is the Einstein tensor, $\xi$ is a dimensionless coupling parameter, and $V(\phi)$ is the potential of the inflaton field. The action given in Eq. (\ref{action}) belongs to a class of the most general scalar-tensor theories having second-order equations of motion \cite{Deffayet2011,Kobayashi2011}. The Lagrangian of such general Horndeski's theories contains the term $G_5(\phi,X)G^{\mu\nu}(\nabla_\mu\nabla_\nu\phi)$, where $G_5$ is a generic function of $\phi$ and $X\equiv-g^{\mu\nu}\partial_\mu\phi\partial_\nu\phi/2$. By choosing the function $G_5 = -\kappa^2\Theta(\phi)/2$, the $\xi$ related term in Eq. (\ref{action}) can be recovered from the Horndeski's Lagrangian after integration by parts. Here the coupling parameter $\xi$ is defined to be $\xi\equiv d\Theta/d\phi$. The case of $\xi=\mathrm{constant}$ that is considered, such as in \cite{Germani2010}, is a special one of $\Theta(\phi)\propto\phi$, which can be used to reconcile steep potentials such as $V\propto\phi^4$ with the current observations of CMB through the mechanism of the gravitationally enhanced friction \cite{Tsujikawa2012}.
But in this paper, we consider a more general case, in which the derivative of $\Theta(\phi)$ with respect to $\phi$ is nontrival, namely $\xi=\theta(\phi)$ is a function of the inflaton $\phi$. What follows are the basic equations in such a general case.

We work in the spatially flat Friedmann-Robertson-Walker background, under which the spacetime line element is written as
\begin{align}\label{metric}
ds^2=-dt^2+a(t)^2 d{\bf x}^2\;.
\end{align}
Then, from the action in Eq. (\ref{action}) with $\xi=\theta(\phi)$, we can obtain the following equations
\begin{align}\label{EOM1}
3H^2=\kappa^{2}\left[\frac{1}{2}\bigg(1+9\kappa^2 \theta(\phi)H^2\bigg)\dot\phi^2+V(\phi)\right]\;,
\end{align}
\begin{align}\label{EOM2}
-2\dot H=\kappa^2\left[\bigg(1+3\kappa^2\theta(\phi)H^2-\kappa^2\theta(\phi)\dot H\bigg)\dot\phi^2-\kappa^2\theta_{,\phi}H\dot \phi^3- 2\kappa^2\theta(\phi)H\dot\phi\ddot\phi\right]\;,
\end{align}

\begin{align}\label{EOM3}
\bigg(1+3\kappa^2 \theta(\phi)H^2\bigg)\ddot\phi + \bigg[1+\kappa^2\theta(\phi)\bigg(2\dot H+3H^2\bigg)\bigg]3H\dot\phi+\frac{3}{2}\kappa^2\theta_{,\phi}H^2\dot\phi^2+V_{,\phi}=0\;,
\end{align}
where $\theta_{,\phi}=d\theta/d\phi$ and $V_{,\phi}=dV/d\phi$.
The slow-roll parameters are defined as 
\begin{align}
\begin{split}\label{SLP}
&\epsilon = -\frac{\dot H}{H^2}\;,\qquad\qquad\;\; \delta_\phi=\frac{\ddot\phi}{H\dot\phi}\;, \\
&\delta_X=\frac{\kappa^2\dot\phi^2}{2H^2}\;,\qquad\qquad \delta_D=\frac{\kappa^4\theta\dot\phi^2}{4}\;.
\end{split}
\end{align}
Since the slow-roll inflation is characterized by
$\{\epsilon,|\delta_\phi|,\delta_X,\delta_D\}\ll 1$, Eqs. (\ref{EOM1})--(\ref{EOM3}) can be approximately  written as
\begin{align}\label{AEOM1}
3H^2\simeq \kappa^2 V(\phi)\;,
\end{align}
\begin{align}\label{AEOM2}
-2\dot H \simeq \kappa^2\left(\mathcal{A}\dot\phi^2-\kappa^2\theta_{,\phi}H\dot\phi^3\right)\;,
\end{align}
\begin{align}\label{AEOM3}
3H \mathcal{A}\dot\phi + \frac{3}{2}\kappa^2\theta_{,\phi}H^2\dot\phi^2+V_{,\phi}\simeq0\;,
\end{align}
where
\begin{align}
\mathcal{A}=1+3\kappa^2\theta(\phi)H^2\;.
\end{align}
The role of the $\theta_{,\phi}$ terms in Eqs. (\ref{AEOM2}) and (\ref{AEOM3}) relies on  the concrete functional form of $\theta(\phi)$. 
For simplicity, we assume that the $\theta_{,\phi}$ terms are negligible during the slow-roll stage, namely
\begin{align}\label{assumption}
|\kappa^2\theta_{,\phi}H\dot\phi|\ll\mathcal{A}\;.
\end{align}
Applying this additional condition, Eqs. (\ref{AEOM2}) and (\ref{AEOM3}) can be further reduced to
\begin{align}\label{FAEOM2}
-2\dot H \simeq \kappa^2\mathcal{A}\dot\phi^2\;,
\end{align}
\begin{align}\label{FAEOM3}
3H \mathcal{A}\dot\phi+V_{,\phi}\simeq0\;.
\end{align}
Using Eqs. (\ref{AEOM1}), (\ref{FAEOM2}) and (\ref{FAEOM3}), we have
\begin{align}\label{epsilon}
\epsilon \simeq \delta_X + 6\delta_D\simeq \frac{\epsilon_V}{\mathcal{A}}\;,
\end{align}
where
\begin{align}
\epsilon_V = \frac{1}{2\kappa^2}\left(\frac{V_{,\phi}}{V}\right)^2\;.
\end{align}
When $\mathcal{A}\simeq1$, $\epsilon\simeq\epsilon_V$ and  one obtains the standard slow-roll inflation. If $\mathcal{A}\gg 1$, $\epsilon\ll \epsilon_V$ due to the high friction, and hence the inflaton rolls more slowly than that in the standard slow-roll inflation. With this feature in mind,  let us  assume that $\mathcal{A}$ has an evolution from $\mathcal{A}\simeq1$ to $\mathcal{A}\gg1$ for a special $\theta(\phi)$, which will decelerate the inflaton and in turn result in the decrease of the slow-roll parameter $\epsilon$. Thus, the amplification of curvature perturbations can be expected in this case. To test this idea, we need to get the power spectrum of the curvature perturbations.

From the action given in Eq. (\ref{action}), one can derive directly the second-order action for the curvature perturbation $\mathcal{R}$~\cite{Kobayashi2011,Tsujikawa2012} 
\begin{align}\label{sec_action}
\mathcal{S}_s^{(2)}=\int dtd^3x a^3Q_s\left[\mathcal{\dot R}^2-
\frac{c_s^2}{a^2}(\partial\mathcal{R})^2\right]\;, 
\end{align}
where
\begin{align}\label{Qs}
Q_s=\frac{w_1(4w_1w_3+9w_2^2)}{3w_2^2}\;,
\end{align}
\begin{align}\label{Cs2}
c_s^2=\frac{3(2w_1^2w_2H-w_2^2w_4+4w_1\dot w_1w_2-2w_1^2\dot w_2)}{w_1(4w_1w_3+9w_2^2)}\;,
\end{align}
and
\begin{align}\label{w}
\begin{split}
&w_1=M_\mathrm{pl}^2(1-2\delta_D)\;,\\
&w_2=2H M_\mathrm{pl}^2(1-6\delta_D)\;,\\
&w_3=-3H^2M_\mathrm{pl}^2(3-\delta_X-36\delta_D)\;,\\
&w_4=M_\mathrm{pl}^2(1+2\delta_D)\;.
\end{split}
\end{align}
It is worth noting that  although the coupling parameter $\xi$ is generalized  to be a function of the inflaton $\phi$, the result given in Eqs.~(\ref{sec_action}-\ref{w}) coincides with the analogous one obtained  in~\cite{Tsujikawa2012}.
In order to avoid the appearance of ghosts and Laplacian
instabilities, we impose the conditions $\{Q_s,c_s^2\}>0$, which can be used to restrict the functional form of $\theta(\phi)$.
At the time when the comoving wave number $k$ exits the horizon [$c_sk=aH$], the power spectrum of the curvature perturbation $\mathcal{R}$ is calculated as \cite{Tsujikawa2012,Kobayashi2011}
\begin{align}
\mathcal{P}_\mathcal{R} = \frac{H^2}{8\pi^2Q_s c_s^3}\;.
\end{align} 
Using Eqs. (\ref{AEOM1}), (\ref{FAEOM2}) and (\ref{FAEOM3}), $\mathcal{P}_\mathcal{R}$ can be approximately expressed as
\begin{align}\label{power}
\mathcal{P}_\mathcal{R} \simeq \frac{V^3}{12\pi^2M_\mathrm{pl}^6V_{,\phi}^2}\left(1+\theta(\phi)\frac{V}{M_\mathrm{pl}^4}\right)\;,
\end{align}
and the scalar spectral index and the tensor-to-scalar ratio are given, respectively, by \cite{Tsujikawa2012}
\begin{align}\label{ns}
n_s\simeq1-\frac{1}{\mathcal{A}}\left[2\epsilon_V\left(4-\frac{1}{\mathcal{A}}\right)-2\eta_V\right]\;,
\end{align}
\begin{align}\label{r}
r\simeq\frac{16\epsilon_V}{\mathcal{A}}\;,
\end{align}
where
\begin{align}
\eta_V=\frac{M_\mathrm{pl}^2}{V}\frac{d^2V}{d\phi^2}\;.
\end{align}
The Planck 2018 results \cite{2Planck2018} give the following constraints on the amplitude of the power spectrum, the scalar spectral index and the tensor-to-scalar ratio at the CMB scale $k_\ast=0.05\;\mathrm{Mpc}^{-1}$,
\begin{align}\label{constraints}
\begin{split}
\ln{(10^{10}\mathcal{P}_\mathcal{R})}=3.044\pm0.014 \qquad &(68\%\;\mathrm{C.L.}),\\
n_s=0.9649\pm0.0042 \qquad &(68\%\;\mathrm{C.L.}), \\
r<0.07 \qquad&(95\%\;\mathrm{C.L.}).
\end{split}
\end{align}

One can see from Eq. (\ref{power}) that the coupling parameter $\theta(\phi)$ with a large peak will lead to a large peak in the power spectrum.
Based on this idea, we consider the following special functional form of $\theta(\phi)$, 
\begin{align}\label{F}
\theta(\phi)=\frac{\omega}{\sqrt{\kappa^2\left(\frac{\phi-\phi_c}{\sigma}\right)^2+1}}\;,
\end{align}
where $\omega$ and $\phi_c$ are the peak height and position, respectively, and $\sigma$ describes the smoothing scale around $\phi=\phi_c$. Such a coupling form  can be obtained when $\Theta(\phi)=\kappa^{-1}\sigma\omega\ln[\kappa(\phi-\phi_c)/\sigma+\sqrt{\kappa^2(\phi-\phi_c)^2/\sigma^2+1}]$.
In this paper, we take into account a simple monomial scalar potential,
\begin{align}\label{V}
V(\phi)=\lambda M_\mathrm{pl}^{4-p}|\phi|^p\;,
\end{align}
with a fractional power $p=2/5$ \cite{Silverstein2008}. Combining Eqs. (\ref{F}) and (\ref{V}), the power spectrum in Eq. (\ref{power}) can be written as
\begin{align}\label{power1}
\mathcal{P}_\mathcal{R} \simeq \frac{\lambda}{12\pi^2 p^2} \bigg |\frac{\phi}{M_\mathrm{pl}} \bigg|^{2+p}\left(1+ \frac{\omega\lambda}{\sqrt{\kappa^2\left(\frac{\phi-\phi_c}{\sigma}\right)^2+1}}\bigg |\frac{\phi}{M_\mathrm{pl}} \bigg|^p\right)\;.
\end{align}
Now we do a simple estimation of the amplification of the curvature perturbations. First of all,  $\phi_\ast$ is introduced as the inflaton value at the time when $k_\ast$ exits the horizon. Then, we assume that 
\begin{align}\label{condition}
\begin{split}
\omega\lambda\gg1\;, \qquad |\phi_\ast-\phi_c|\gg\omega\lambda\sigma M_\mathrm{pl}\;.
\end{split} 
\end{align}
Finally, we have
\begin{align}\label{peak}
\mathcal{P}_\mathcal{R}|_{\phi=\phi_c}\simeq \omega\lambda \bigg|\frac{\phi_c}{\phi_\ast}\bigg|^{2+p}\bigg|\frac{\phi_c}{M_\mathrm{pl}}\bigg|^p \mathcal{P}_\mathcal{R}|_{\phi=\phi_\ast}\;.
\end{align}
Obviously, the power spectrum at $\phi=\phi_c$ is amplified by $\mathcal{O}(\omega\lambda$) orders of magnitude relative to that at $\phi=\phi_\ast$, whose amplitude is constrained as $2.10\times10^{-9}$ by the CMB observations. Therefore, the power spectrum will have a peak of $\mathcal{O}(10^{-2})$ on the scale corresponding to $\phi=\phi_c$ in the case of $\omega\lambda\sim\mathcal{O}(10^7)$, which could lead to the formation of PBHs. In next section, we will study the concrete examples by numerical methods.

\section{Numerical results}
\label{sec4}

\begin{table}
\caption{The successful parameter sets for producing the PBHs with mass around $\mathcal{O}(10)M_\odot$ (\emph {Case 1}), $\mathcal{O}(10^{-5})M_\odot$ (\emph {Case 2}) and $\mathcal{O}(10^{-12})M_\odot$ (\emph {Case 3}).}
\begin{tabular}{>{\centering}p{2cm}>{\centering}p{2cm}>{\centering}p{2cm}>{\centering}p{2cm}}
\hline
\hline 
$\#$ & $\phi_{c}/M_\mathrm{pl}$ & $\omega\lambda$ & $\sigma$  \tabularnewline
\hline 

\emph {Case 1} & $4.63$ & $1.33\times10^7$ & $2.6\times10^{-9}$  \tabularnewline

\emph {Case 2} & $3.9$ & $1.53\times10^7$ & $3\times10^{-9}$   \tabularnewline

\emph {Case 3} & $3.3$ & $1.978\times10^7$ & $3.4\times10^{-9}$  \tabularnewline
\hline 
\end{tabular}
\label{table1}
\end{table} 

Respecting the conditions in Eq. (\ref{condition}), we consider the three concrete sets of parameters shown in Table \ref{table1}. For these parameter sets, we find that the effect of the nonminimal derivative coupling has almost disappeared at the end of inflation. The value of $\phi$ at the end of inflation in these cases is the same as that in the standard slow-roll inflation. Thus, we can get the end value $\phi_f\simeq0.28M_\mathrm{pl}$ by solving $\epsilon_V\simeq1$ for these cases. 
We set the \textit{e}-folding number from the time when $k_\ast$ exits horizon to the end of the inflation as $N_\ast=60$ for  case 1 and 2, and as $N_\ast=65$ for case 3.
Table \ref{table2} gives the corresponding derived cosmological parameters and the quantities associated with the produced PBHs for these cases. 
Next, taking case 1 as an example, we study the inflationary dynamics of this model by solving the equations numerically. 

\begin{table}
\caption{Results for the three cases of table \ref{table1}. $\mathcal{P_{R}}^{\mathrm{peak}}$ and $f_{\mathrm{PBH}}^{\mathrm{peak}}$ are the peak values of the power spectra of the curvature perturbations and the mass spectra of PBHs, respectively. $M_{\mathrm{PBH}}^{\mathrm{peak}}$ is the PBH mass corresponding to $f_{\mathrm{PBH}}^{\mathrm{peak}}$.}
\begin{tabular}{>{\centering}p{1.3cm}>{\centering}p{1.3cm}>{\centering}p{2cm}>{\centering}p{1.8cm}>{\centering}p{1.8cm}>{\centering}p{1.8cm}>{\centering}p{1.8cm}>{\centering}p{1.8cm}>{\centering}p{1.8cm}}
\hline 
\hline 
$\#$ & $\phi_{\ast}/M_\mathrm{pl}$ & $\lambda$ & $n_{s}$ & $r$ & $\mathcal{P_{R}}^{\mathrm{peak}}$ & $M_{\mathrm{PBH}}^{\mathrm{peak}}/M_\odot$ & $f_{\mathrm{PBH}}^{\mathrm{peak}}$ & $\Omega_{\mathrm{PBH}}/\Omega_{\mathrm{DM}}$\tabularnewline
\hline 

\emph {Case 1} & $4.99$ & $7.09\times10^{-10}$ & $0.9666$ & $0.0431$ & $0.0473$ & $23.5$ & $1.88\times10^{-3}$ & $1.95\times10^{-3}$\tabularnewline

\emph {Case 2} & $4.83$ & $8.23\times10^{-10}$ & $0.9618$ & $0.0497$ & $0.0386$ & $9.02\times10^{-6}$ & $0.0452$  & $0.044$ \tabularnewline

\emph {Case 3} & $4.77$ & $8.52\times10^{-10}$ & $0.9607$ & $0.0512$ & $0.0312$ & $8.1\times10^{-13}$ &$0.977$ & $ 0.972$\tabularnewline
\hline 
\end{tabular}
\label{table2}
\end{table}

\begin{figure}
\centering
\subfigure{\label{fig1a}}{\includegraphics[width=0.48\textwidth ]{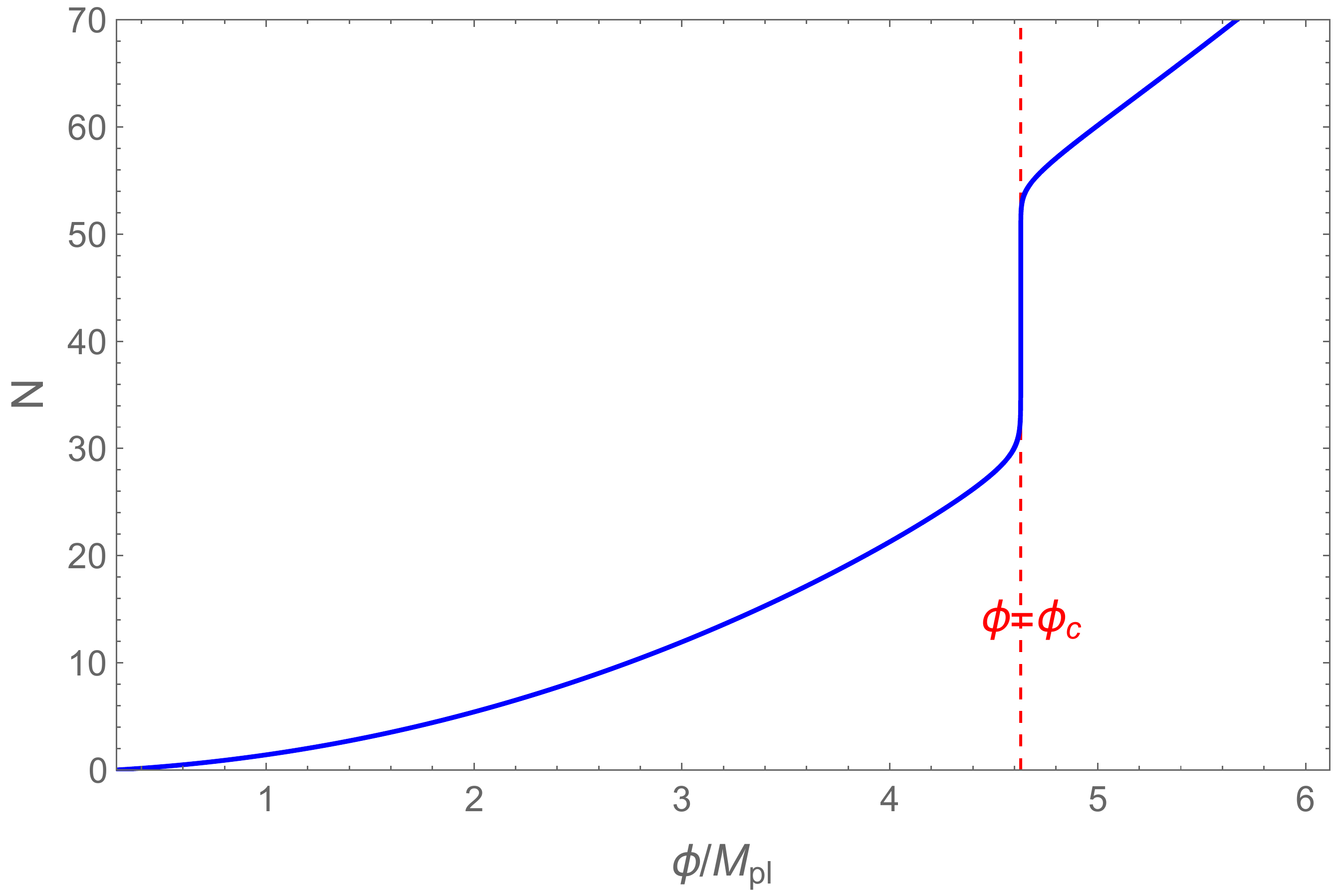}}
\subfigure{\label{fig1b}}{\includegraphics[width=0.48\textwidth ]{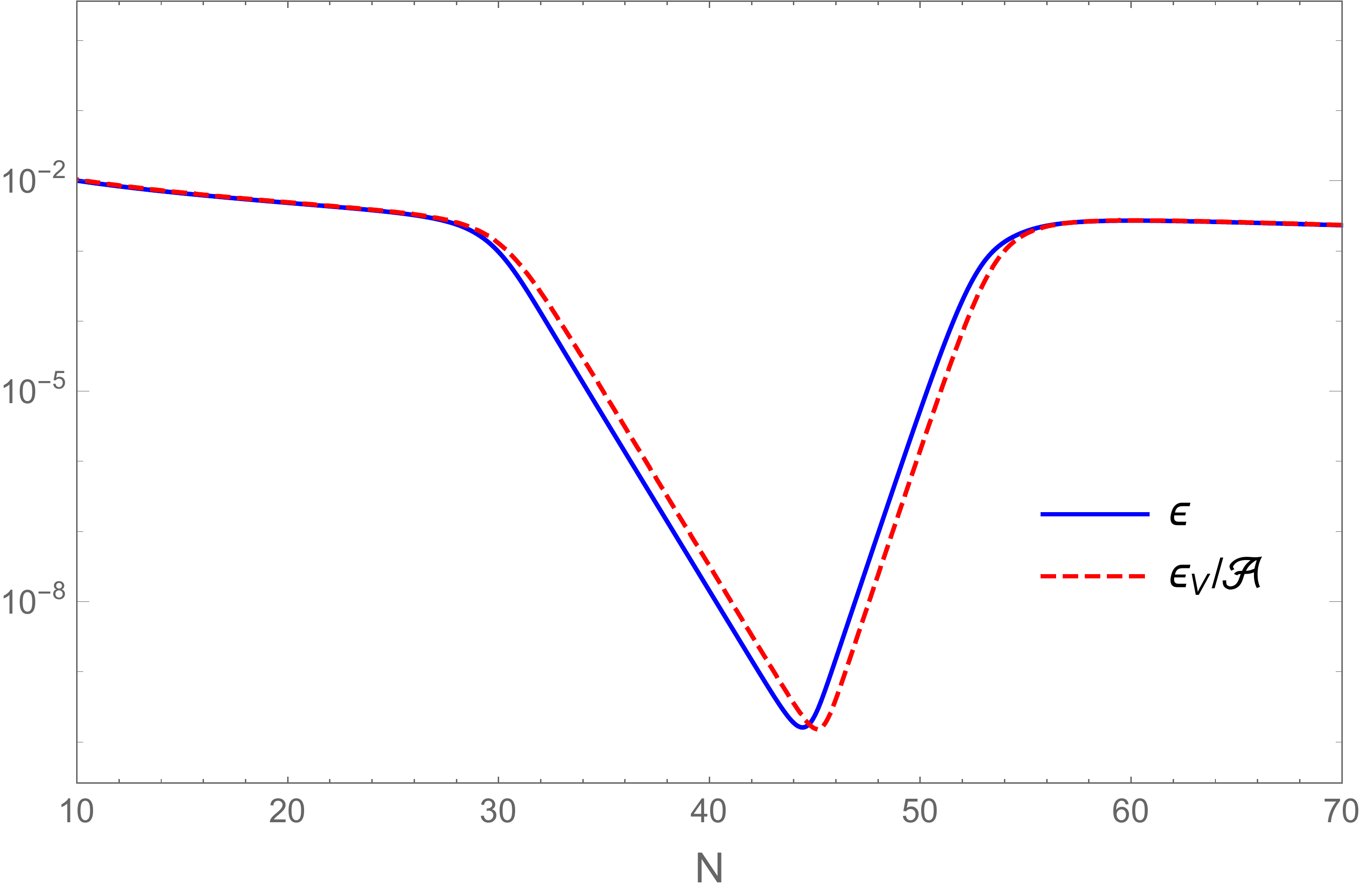}}
\caption{\label{fig1} (a) \textit{Left-hand plot}: The evolution of the \textit{e}-folding number $N\equiv \ln[a(\phi_f)/a(\phi)]$ as a function of $\phi$ in case 1. (b) \textit{Right-hand plot}: The evolutions of $\epsilon$ (solid line) and $\epsilon_V/\mathcal{A}$ (dashed line) versus $N$ in case 1.}
\end{figure}

\begin{figure}
\centering
\includegraphics[width=0.8\textwidth ]{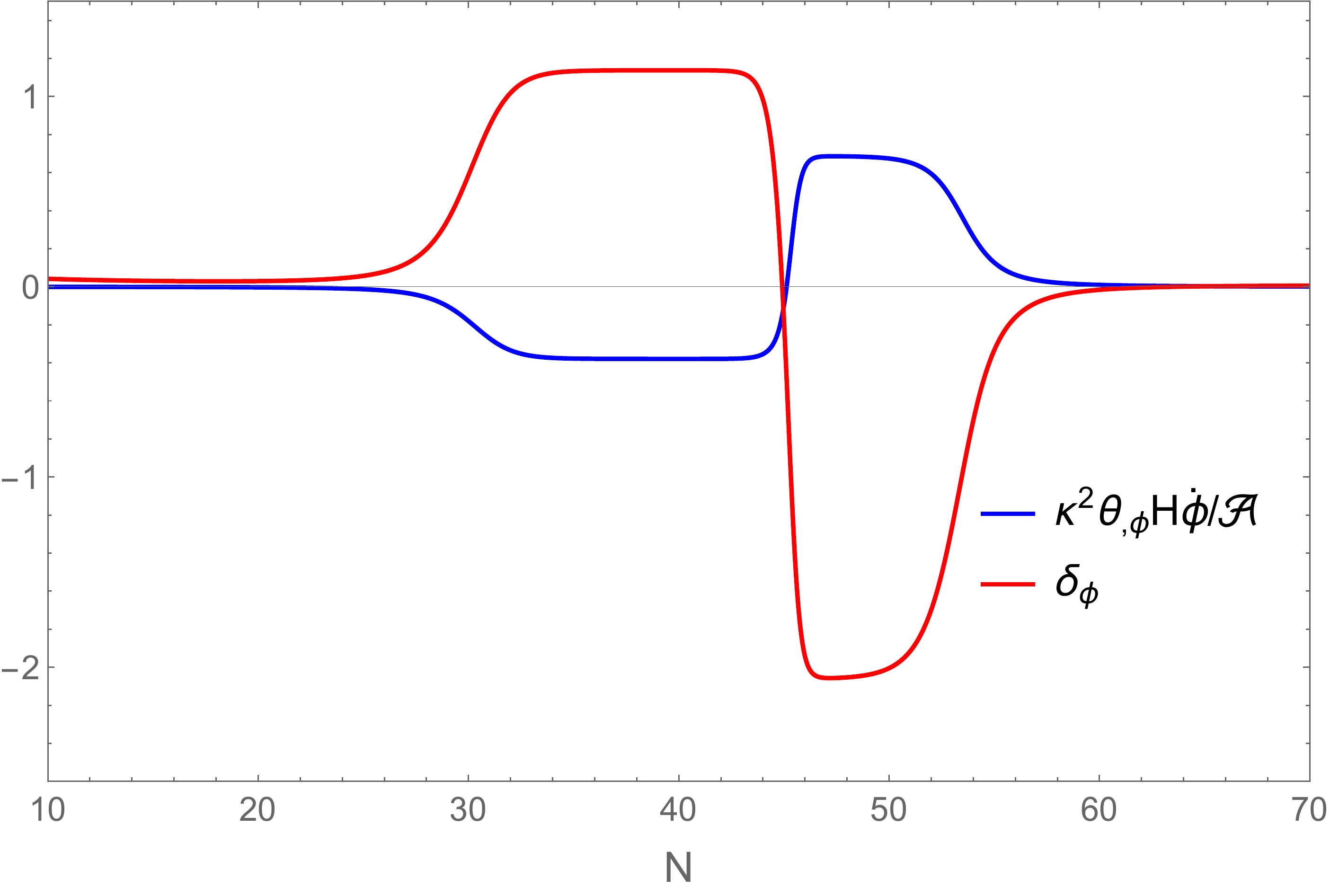}
\caption{\label{fig2} The evolutions of $\kappa^2\theta_{,\phi}H\dot\phi/\mathcal{A}$ (blue line) and the slow-roll parameter $\delta_\phi$ (red line) as a function of $N$ in case 1.} 
\end{figure}

In Fig. \ref{fig1a}, we plot the evolution of the number of \textit{e}-folds $N\equiv \ln[a(\phi_f)/a(\phi)]$ as a function of $\phi$. One can see that the inflaton almost stops rolling at around $\phi=\phi_c$ due to the high friction, and it takes the inflaton about 20 e-folds to cross $\phi_c$. This shows that the Universe experiences a period of ultra-slow-roll inflation corresponding to $30<N<54$. Figure \ref{fig1b} shows the evolutions of $\epsilon$ and $\epsilon_V/\mathcal{A}$ as a function of $N$. When $N>54$, it can be seen that $\epsilon$ is almost coincident with $\epsilon_V/\mathcal{A}$, which implies that the approximate equations (\ref{AEOM1}), (\ref{FAEOM2}) and (\ref{FAEOM3}) are valid at this phase. Thus, after getting the inflaton value $\phi_\ast=4.99M_\mathrm{pl}$ corresponding to $N_\ast=60$, the scalar spectral index and the tensor-to-scalar ratio at the pivot scale $k_\ast$ can be calculated by using Eqs. (\ref{ns}) and (\ref{r}) to be  $n_s=0.9666$ and $r=0.0431$, which are compatible with the current observational constraints in Eq. (\ref{constraints}).  Combining Eq. (\ref{power1}) with $\mathcal{P}_\mathcal{R}|_{\phi=\phi_\ast}\simeq 2.10\times10^{-9}$, we obtain $\lambda\simeq 7.09\times 10^{-10}$. When $30<N<54$, one can see that the slow-roll parameter $\epsilon$ decreases by seven orders of magnitude, which is a result of the ultra-slow-roll inflation. Accordingly, the curvature perturbations will be amplified by seven orders of magnitude. However, $\epsilon_V/\mathcal{A}$ deviates from the slow-roll parameter $\epsilon$ during this period. This is because  one of the slow-roll conditions $|\delta_\phi| \ll1$ and the additional condition in Eq. (\ref{assumption}) are violated when $30<N<54$, as shown in Fig. \ref{fig2}, which shows the evolutions of $\kappa^2\theta_{,\phi}H\dot\phi/\mathcal{A}$ and the slow-roll parameter $\delta_\phi$ as a function of $N$.  As a result, the formula (\ref{power}) is just a rough estimation of the enhanced power spectrum, which is obtained on the premise that all slow-roll conditions and the condition in Eq. (\ref{assumption}) are valid.

\begin{figure}
\centering
\includegraphics[width=0.8\textwidth ]{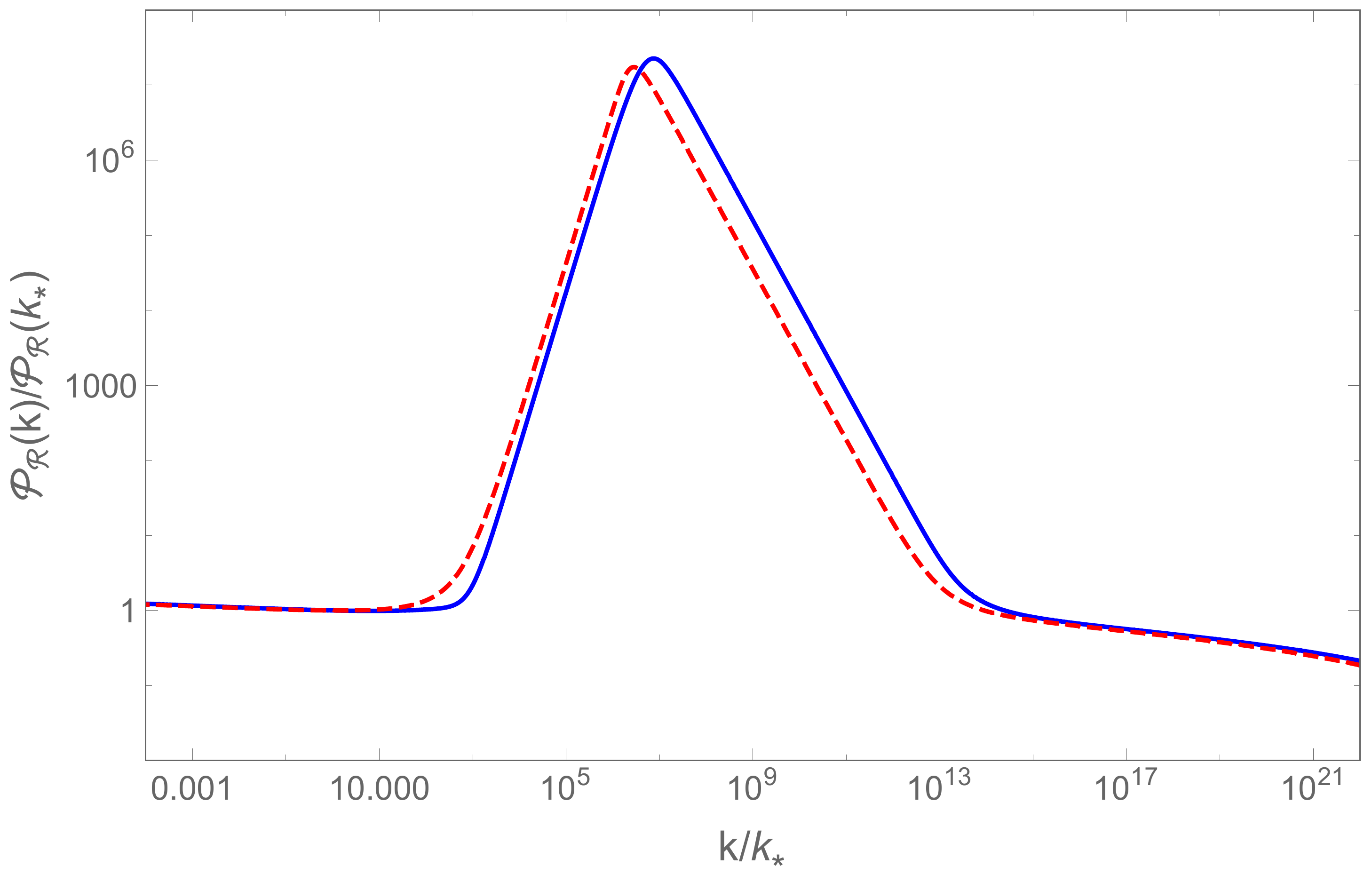}
\caption{\label{fig3} The power spectra $\mathcal{P_R}(k)/\mathcal{P_R}(k_\ast)$ as a function of $k/k_\ast$ in case 1. The power spectrum is calculated by solving the Mukhanov-Sasaki equation (blue line) and using the approximate solution in Eq. (\ref{power}) (red line).   } 
\end{figure}

In order to obtain the exact power spectrum of the curvature perturbations, we need to resort to numerical solution of  the Mukhanov-Sasaki equation: 
\begin{align}\label{MS}
u_k^{''} + \left(c_s^2k^2-\frac{z^{''}}{z}\right)u_k=0\;,
\end{align}
which is obtained  by varying the action (\ref{sec_action}) with respect to $u$, where  $z\equiv\sqrt{2Q_s}a$ and $u\equiv z\mathcal{R}$ are  the new variables, and the prime denotes the derivative with respect to the conformal time $\eta\equiv\int a^{-1}dt$. With these new variables, the exact power spectrum of the curvature perturbations has the form $\mathcal{P_R}(k)=(2\pi^2)^{-1}k^3|u_k/z|^2$. 
Figure \ref{fig3} compares the power spectra of the curvature perturbations from the approximate solution in Eq. (\ref{power}) and the exact numerical solution of the Mukhanov-Sasaki equation. It can be seen that the the formula (\ref{power}) can reproduce to some extent the qualitative behavior of the actual power spectrum. Although the peak value of the power spectrum from formula (\ref{power}) is only slightly smaller than that of the actual power spectrum, $\beta$ is exponentially sensitive to small variations of the power spectrum, and our calculations show that the predicted abundance of PBHs from the power spectrum in Eq. (\ref{power}) is three orders of magnitude smaller than that from the actual power spectrum. Therefore, it is necessary to obtain the power spectrum of the curvature perturbations by numerically solving the Mukhanov-Sasaki equation. 


\begin{figure}
\centering
\includegraphics[width=0.8\textwidth ]{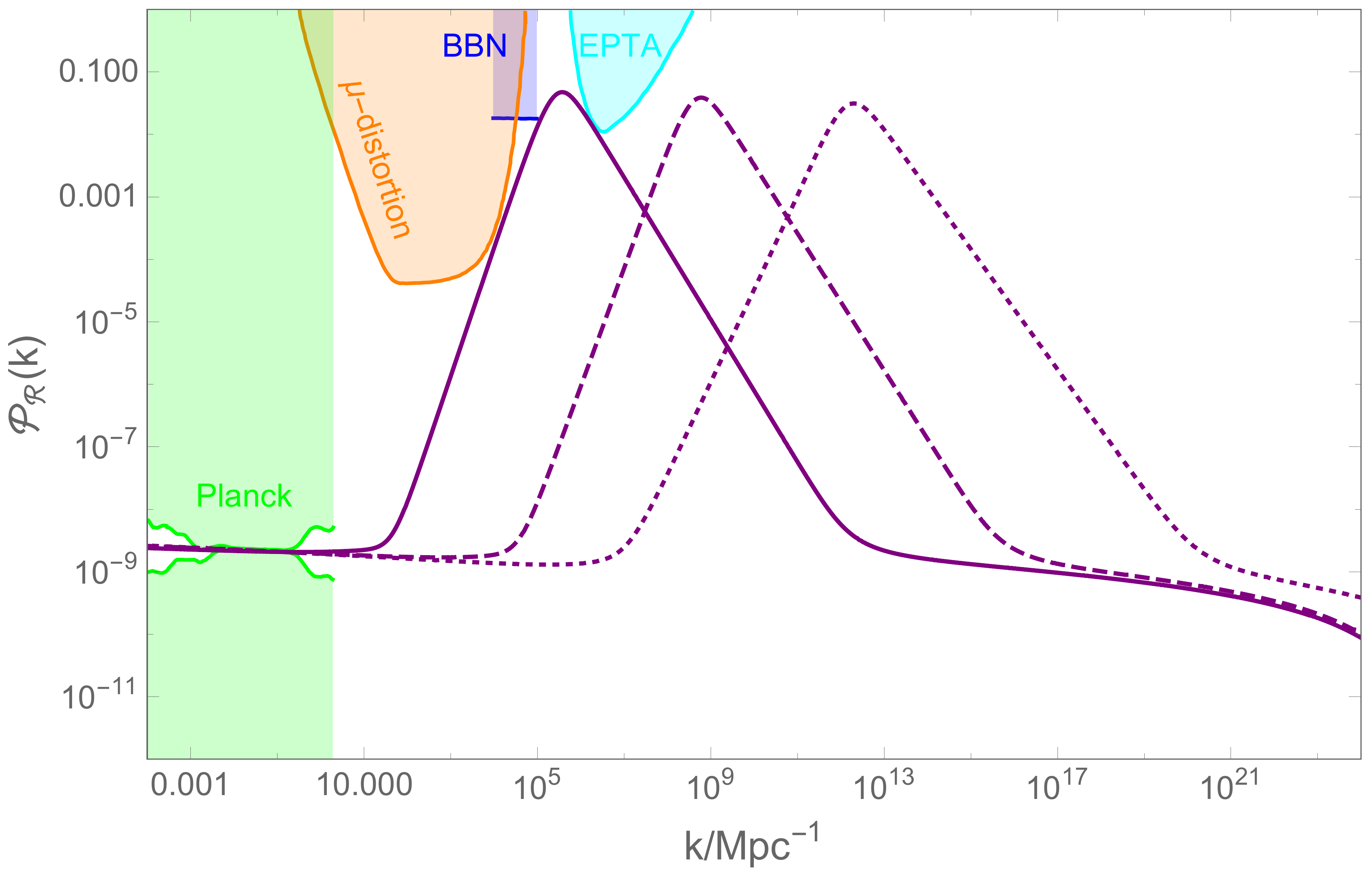}
\caption{\label{fig4} The power spectra of the curvature perturbations obtained by solving the Mukhanov-Sasaki equation as a function of the comoving wave number $k$ in case 1 (solid line), case 2 (dashed line), and case 3 (dotted line). The green-shaded region is excluded by the current CMB observations \cite{2Planck2018}. The orange- and blue-shaded regions are excluded by the $\mu$-distortion of CMB \cite{distortion} and the effect on n-p ratio during big-bang nucleosynthesis (BBN) \cite{BBN}, respectively. The cyan-shaded region shows the constraints on the power spectrum from the current PTA observations \cite{PTA}, which is obtained by parametrizing the power spectrum profile of curvature perturbations as $\mathcal{P_R}\sim \exp [-(\log k-\log k_p)^2/\tilde{\sigma}^2]$ with $\tilde{\sigma}=0.5$. }  
\end{figure}

\begin{figure}
\centering
\includegraphics[width=0.8\textwidth ]{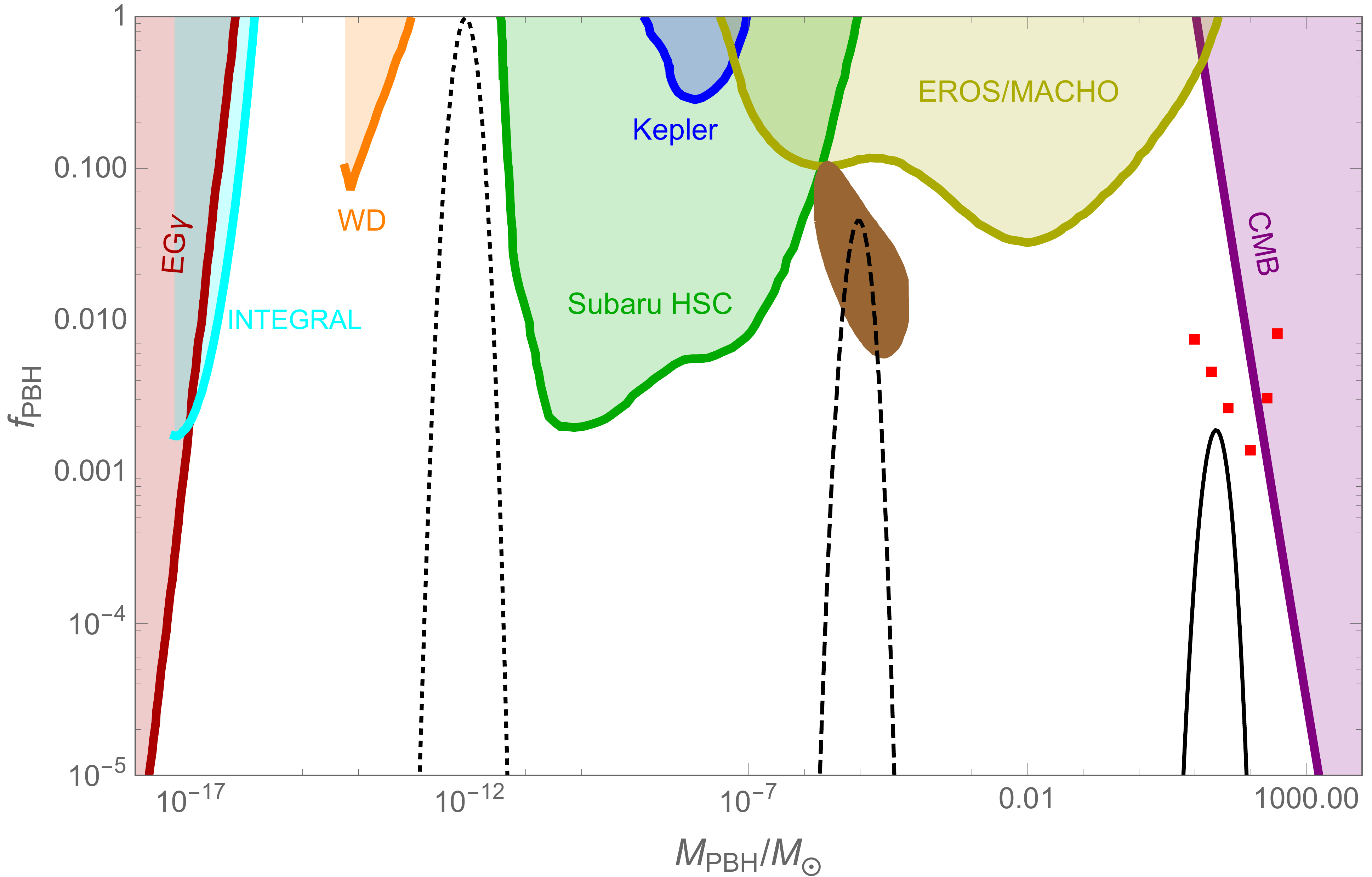}
\caption{\label{fig5} The mass spectra of PBHs for case 1 (solid line), case 2 (dashed line), and case 3 (dotted line). The red points represent the potential upper bounds on the PBH abundance from requiring that the merger rate of PBHs does not exceed the upper limit on the LIGO merger rate \cite{LIGO}.
The allowed PBH abundance from the ultrashort-timescale microlensing events in the OGLE data \cite{OGLE1,OGLE2} is shown in brown-shaded region,
while the other shaded regions show the current observational constraints: extragalactic gamma-rays from PBH evaporation (EG$\gamma$) \cite{EG}, galactic center 511 keV gamma-ray line (INTEGRAL) \cite{INTEGRAL}, white dwarfs explosion (WD) \cite{WD}, microlensing events with Subaru HSC (Subaru HSC) \cite{HSC}, with the Kepler satellite (Kepler) \cite{Kepler}, with EROS/MACHO (EROS/MACHO) \cite{EROS}, and accretion constraints from CMB (CMB) \cite{CMB}.
 } 
\end{figure}

In Fig. \ref{fig4}, we show the actual power spectra for the cases of Table \ref{table1} and the existing observational constrains on the power spectrum. One can see  that the power spectra in these cases all remain nearly scale invariant on large scales, which are compatible with the Planck 2018 results. In particular, for case 1, the power spectrum has a sharp peak around $10^{5}$--$10^{6}\mathrm{Mpc}^{-1}$, which just meets the constrains from CMB $\mu$-distortion, BBN, and PTA observations. The predicted mass spectra of PBHs are plotted in Fig. \ref{fig5}, in which the current observational constraints on the PBH abundance are also shown. In case 1, the resulting mass spectrum of PBHs has a sharp peak at $23.5M_\odot$, and its height is $1.88\times10^{-3}$, which can explain the LIGO events and satisfy the constrains from the upper limit on the LIGO merger rate. In case 2, the peak of the PBH mass spectrum, whose height and position are $0.0452$ and $9.02\times10^{-6}M_\odot$ respectively, locates in the inferred region of the PBH abundance by the ultrashort-timescale microlensing events in OGLE data. Thus, the formed PBHs in case 2 can be taken as a source of these microlensing events. The resulting PBHs with mass around $\mathcal{O}(10^{-12})M_\odot$ comprising $97.2\%$ of the total DM can be taken as a main component of DM in case 3.

\section{conclusions and discussions}
\label{sec5}
In this paper, we have investigated the possibility of the PBH formation in a single-field model of inflation with a field derivative coupling to the Einstein tensor. By introducing the coupling parameter as a special function of the inflaton field, which can be realized in a class of the most general scalar-tensor theories \cite{Kobayashi2011,Deffayet2011}, we succeeded in realizing a period of ultra-slow-roll inflation resulted from the gravitational enhanced friction. Consequently, the curvature perturbations have a sharp peak, which is large enough to produce a sizeable amount of PBHs. With this mechanism, we calculated the power spectra of the curvature perturbations in three parameter sets with a fractional power-law potential $V\propto\phi^{2/5}$. And then we obtained the sharp mass spectra of PBHs at $\mathcal{O}(10)M_\odot$, $\mathcal{O}(10^{-5})M_\odot$ and $\mathcal{O}(10^{-12})M_\odot$, which can explain the LIGO events, the ultrashort-timescale microlensing events in OGLE data, and the most of DM, respectively. Note that, with just three parameters, we can obtain a power spectrum of the curvature perturbations, which
has a large enough  peak on the small scales while satisfying the current observational constraints on the large scales.
Another nice property is that the PBH mass spectrum in which we are interested can be easily obtained through fine-tuning two of these three parameters.
Finally, we point out that  large-amplitude curvature perturbations not only collapse to form PBHs, but also induce the GWs through their second-order effects, as discussed in \cite{Kohri2018,Pi2019,Wang2019,Inomata2019,Liu2019}. There is also no exception about the production of GWs in our model, but we left this interesting issue for future work.

\begin{acknowledgments}
We appreciate very much the insightful comments and helpful suggestions by anonymous referees, and thank Jing Liu for fruitful discussions. This work was supported by the National Natural Science Foundation of China under Grants No. 11775077, No. 11435006, and No. 11690034, and by the Science and Technology Innovation Plan of Hunan province under Grant No. 2017XK2019.
\end{acknowledgments}

\end{document}